\documentclass[letter,11pt,DIV=12,abstract=true,numbers=noenddot,titlepage=false,twocolumn=false,draft=false]{scrartcl}
\pdfoutput=1

\makeatletter
\DeclareOldFontCommand{\rm}{\normalfont\rmfamily}{\mathrm}
\DeclareOldFontCommand{\sf}{\normalfont\sffamily}{\mathsf}
\DeclareOldFontCommand{\tt}{\normalfont\ttfamily}{\mathtt}
\DeclareOldFontCommand{\bf}{\normalfont\bfseries}{\mathbf}
\DeclareOldFontCommand{\it}{\normalfont\itshape}{\mathit}
\DeclareOldFontCommand{\sl}{\normalfont\slshape}{\@nomath\sl}

\usepackage[usenames,dvipsnames]{xcolor}
  \definecolor{hgreen}{rgb}{0,.3,0}
  \definecolor{hred}{rgb}{.3,0,0}
  \definecolor{hblue}{rgb}{0,0,.3}
  \definecolor{LightGray}{gray}{0.95}
  \definecolor{gray}{gray}{0.6}

\usepackage[
	    colorlinks=true,
	    linkcolor=hblue,
	    citecolor=hgreen,
	    filecolor=hblue,
	    urlcolor=hred
	    ]{hyperref}
\usepackage{mathrsfs}
\usepackage{soul}
\usepackage[intlimits]{amsmath}
\usepackage{amssymb}
\usepackage{slashed}
\usepackage{booktabs}
\usepackage[titletoc,title]{appendix}
\usepackage[affil-it]{authblk}
\usepackage{libertine}
\usepackage[numbers,sort&compress]{natbib}
\usepackage{graphicx}
\usepackage{tensor}

\usepackage[protrusion=true,expansion,kerning=true,tracking=true,final]{microtype}

\allowdisplaybreaks
\setcapindent{1em}
\setkomafont{captionlabel}{\bfseries}
\setkomafont{caption}{\itshape}

\newcommand{\Nf}{N_{\!f}}
\newcommand{\Lag}{\mathscr{L}}
\newcommand{\muew}{\mu_{\text{ew}}}

\newcommand{\q}{{q_h}}
\newcommand{\qb}{{\bar{q}_h}}
\newcommand{\GF}{G_{\mathrm{F}}}

\newcommand{\alphas}[1]{\alpha_{s,\text{#1fl}}}
\newcommand{\alphast}[1]{\tilde{\alpha}_{s,\text{#1fl}}}

\newcommand{\kkappa}[1]{\kappa_{\text{#1fl}}}

\begin{document}
\titlehead{\hfill DO-TH 23/07}

\renewcommand\Authands{, }

\title{\boldmath 
  A Precise Electron EDM Constraint on CP-odd Heavy-Quark Yukawas
}

\date{\today}
\author[a]{Joachim Brod%
        \thanks{\texttt{joachim.brod@uc.edu}}}
\author[b]{Zachary Polonsky%
        \thanks{\texttt{zach.polonsky@physik.uzh.ch}}}
\author[c]{Emmanuel Stamou%
        \thanks{\texttt{emmanuel.stamou@tu-dortmund.de}}}
	\affil[a]{{\large Department of Physics, University of Cincinnati, Cincinnati, OH 45221, USA}}
	\affil[b]{{\large Physik-Institut, Universit\"at Z\"urich, CH-8057 Z\"urich, Switzerland}}
	\affil[c]{{\large Fakult\"at Physik, TU Dortmund, D-44221 Dortmund, Germany}}

\maketitle

\begin{abstract}
  CP-odd Higgs couplings to bottom and charm quarks arise in many
  extensions of the standard model and are of potential interest for
  electroweak baryogenesis. These couplings induce a contribution to
  the electron EDM. The experimental limit on the latter then leads to
  a strong bound on the CP-odd Higgs couplings. We point out that this
  bound receives large QCD corrections, even though it arises from a
  leptonic observable. We calculate the contribution of CP-odd Higgs
  couplings to the bottom and charm quarks in renormalisation-group
  improved perturbation theory at next-to-leading order in the strong
  interaction, thereby reducing the uncertainty to a few percent.
\end{abstract}
%\pdfbookmark[1]{Table of Contents}{tableofcontents}
\setcounter{page}{1}
%\tableofcontents

\section{Introduction\label{sec:introduction}}

The precise determination of the Yukawa couplings of the Higgs boson
to all fermions has been a focus of particle physics since the
discovery of the Higgs boson in 2012~\cite{ATLAS:2012yve,
  CMS:2012qbp}. In the Standard Model (SM) all Yukawa couplings are
aligned with the fermion masses and thus real, but multiple extensions
of the SM induce non-trivial phases. This is of particular interest as
these phases (mainly in the Yukawa couplings to the third fermion
generation) play a leading role in models of electroweak
baryogenesis~\cite{deVries:2017ncy, DeVries:2018aul,
  Fuchs:2020uoc}. In this article, we focus on the bottom- and
charm-quark Yukawa couplings.

It is well-known that the present bounds on the electric dipole moment
(EDM) of the electron place strong constraints on CP-violating phases
in the quark Yukawa couplings~\cite{Brod:2013cka, Chien:2015xha,
  Egana-Ugrinovic:2018fpy, Brod:2018pli, Bahl:2022yrs, Brod:2022bww}.
However, it is less well-appreciated (and maybe somewhat surprising)
that the heavy-quark contributions to the electron EDM receive large
QCD corrections, leading to a large implicit uncertainty in the
current constraints. In this section, we briefly review the current
situation.  In the remainder of this article, we calculate the leading
logarithmic (LL) and next-to-leading logarithmic (NLL) QCD
corrections, thereby reducing the presently ${\cal O}(1)$ uncertainty
to a few percent.

For the purpose of this work, we assume a modification of the SM
heavy-quark Yukawa couplings of the form\footnote{ This
  parameterisation of pseudoscalar Higgs couplings should be thought
  of as either the dimension-four part of the so-called Higgs
  Effective Field Theory~\cite{Feruglio:1992wf}, the electroweak
  chiral Lagrangian\cite{Buchalla:2013rka} in {\itshape unitarity
    gauge} for the electroweak sector, or as the leading term arising
  from the dimension-six SMEFT operators of the form
  $H^\dagger H \overline{Q}_{L,i}{H} d_{R,i}$ and
  $H^\dagger H \overline{Q}_{L,i}{\tilde{H}} u_{R,i}$, where $H$
  denotes the Higgs doublet in the unbroken phase of electroweak gauge
  symmetry, while $Q_{L,i}$ and $d_{R,i}$/$u_{R,i}$ with $i = 2,3$
  represent the left-handed quark doublet and the right-handed quark
  fields of the second or third generation, respectively. For further
  details see also the discussion in Ref.~\cite{Brod:2022bww}.  }
\begin{equation}\label{eq:LagHqq}
  \Lag_{h\q\q} = - \frac{y^{\text{SM}}_\q}{\sqrt{2}} \kappa_\q
  \qb\left( \cos\phi_\q + i\gamma_5\sin\phi_\q \right)\q\,h\,.
\end{equation}
Here, $\q = b, c$ denotes the bottom- or charm-quark field and $h$ the
physical Higgs field. Moreover,
$y^{\text{SM}}_\q \equiv m_\q e/\sqrt{2}s_wM_W$ is the SM Yukawa, with
$e$ the positron charge, $s_w$ the sine of the weak mixing angle, and
$m_q$ and $M_W$ the heavy-quark and $W$-boson masses,
respectively. The real parameter $\kappa_\q\geq0$ parameterises
modifications to the absolute value of the Yukawa coupling, while the
phase $\phi_\q \in [0,2\pi)$ parameterises CP violation. The SM
corresponds to the values $\kappa_\q=1$ and $\phi_\q=0$.

\begin{figure}[t]
  \centering
  \includegraphics{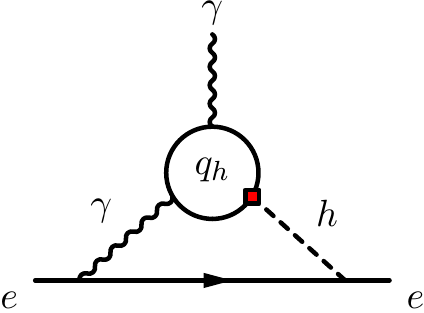}
  \caption{Barr--Zee diagram that contributes to the
	electron EDM from a CP-odd heavy-quark Yukawa coupling with
	the Higgs (red square). \label{fig:barrzee}}
\end{figure}

Virtual heavy quarks with CP-odd Higgs couplings induce an
electron EDM $d_e$, defined by
\begin{equation} \label{eq:LeffN}
\Lag_{\rm eff} = - d_e \, \frac{i}{2} \, \bar e \sigma^{\mu\nu} \gamma_5 e \, F_{\mu\nu} \,,
\end{equation}
with $\sigma^{\mu\nu} \equiv \tfrac{i}{2} [\gamma^\mu, \gamma^\nu]$, via
the well-known Barr--Zee diagrams~\cite{Barr:1990vd}. Calculating the two-loop
Barr--Zee diagrams (see Figure~\ref{fig:barrzee}) 
keeping the non-zero heavy-quark mass and expanding the loop
functions for small quark mass and small external momenta
leads to~\cite{Brod:2013cka}
\begin{equation}\label{eq:dsw}
\begin{split}
d_e & \simeq -12 e Q_e Q_\q^2 \, \frac{\alpha_e}{(4\pi)^3} \sqrt{2}
G_F m_e \, \kappa_\q \sin\phi_\q \, x_\q \left (
\log^2 x_\q + \frac{\pi^2}{3} \right ) \,,
\end{split}
\end{equation}
up to higher orders in the ratio $x_\q \equiv m_\q^2/M_h^2$ (with
$M_h \sim 125\,$GeV the Higgs-boson mass). Here, $\alpha_e$ is the
fine-structure constant, $Q_\q$ is the charge of the heavy quark with
$Q_b = -1/3$ and $Q_c = +2/3$, and $Q_e = -1$ is the electron charge.
Barr--Zee diagrams with an internal $Z$ boson also affect the electron
EDM, however, they lead to a contribution that is suppressed with
respect to those with an internal photon by the small coupling of the
$Z$ boson to electrons~\cite{Brod:2013cka}.

Using the recent bound $|d_e| < 4.1 \times 10^{-30} \, e \, \text{cm}$
(at 90\% CL)~\cite{Roussy:2022cmp}, Eq.~\eqref{eq:dsw} implies the
bounds $\kappa_b|\sin\phi_b| \leq 0.15$ and
$\kappa_c|\sin\phi_c| \leq 0.41$ at the 90\%\,CL for the bottom- and
charm-quark case, respectively.  However, to obtain these bounds,
numerical values for the heavy-quark masses must be chosen. It is not
clear {\em a priori} at which scale the quark masses should be
evaluated; $\mu = M_h$ and $\mu = m_\q$ would be obvious choices. For
the bounds above, we used the values $m_b(M_h)$ and $m_c(M_h)$;
choosing $m_b(m_b)$ and $m_c(m_c)$ instead leads to the significantly
stronger bounds $\kappa_b|\sin\phi_b| \leq 0.086$ and
$\kappa_c|\sin\phi_c| \leq 0.13$ at the 90\%\,CL. The differences
arise from the large QCD running of the quark masses between the two
scales $\mu = M_h$ and $\mu = m_\q$. This indicates that the QCD
corrections to Eq.~\eqref{eq:dsw} are large, even though the electron
EDM is a leptonic observable. By our explicit calculation, we will
show that the predicted value for $d_e$ after resolving the ambiguity
lies somewhere in between the values obtained by using the two scales
$M_h$ and $m_\q$.

We give here a brief outline of the main ideas of the calculation.
First, notice that the result in Eq.~\eqref{eq:dsw}, which was
obtained by a two-loop, fixed-order calculation, is numerically
dominated by the large quadratic logarithms $\log^2 x_\q$. This
logarithmic contribution can be reproduced by the {\em one-loop} QED
renormalisation-group (RG) evolution in an appropriate effective
theory (see Section~\ref{sec:eft}), truncated at order $\alpha_e^2$. The
second term in Eq.~\eqref{eq:dsw}, $\pi^2/3$, has no logarithmic
dependence on the mass ratio and is formally of
next-to-next-to-leading-logarithmic (NNLL) order in RG-improved
perturbation theory. Numerically, this term gives a $\sim 6\%$
correction to the logarithmic contribution to $d_e/e$ in the bottom
case and a $\sim 3.5\%$ correction in the charm case. On the other
hand, we have seen above that the choice of different renormalisation
scales for the quark masses represents a much larger uncertainty, of
the order of $100\%$.

We, thus, conclude that the QCD corrections to the Barr--Zee diagrams
are large, and we expect that these corrections are dominated by
leading QCD logarithms, since the product $\alpha_s \log x_\q$ is
large ($\alpha_s$ denotes the strong coupling constant).  These
logarithms can be reliably calculated using RG-improved perturbation
theory. The result with resummed leading QCD logarithms has the
schematic form
$\alpha_e^2 \log^2 x_\q (1 + \alpha_s \log x_\q + \alpha_s^2 \log^2
x_\q + \ldots)$. Here, the first term reflects the quadratic logarithm
in Eq.~\eqref{eq:dsw}, while all other terms correspond to the leading
logarithms of the diagrams obtained by dressing the Barr-Zee diagrams,
Figure~\ref{fig:barrzee}, with an arbitrary number of gluons.

It is, however, well-known that one can only consistently fix the scale and
scheme dependence of the input parameters by going beyond the LL
approximation. Hence, we will also perform the NLL calculation, reducing the
the uncertainty down to the percent level. This corresponds
roughly to the size of the correction that we have estimated above
using the $\pi^2/3$ term in the fixed-order result, which can be
viewed as part of the NNLL order result in RG-improved perturbation
theory. This term is thus not included in our NLL calculation.

Our calculation will show that the QCD perturbation series converges
well, as might be expected for a leptonic observable. In addition, we
emphasize that this is a complete calculation -- no further hadronic
input (such as the lattice matrix elements required for hadronic EDMs)
is needed. The final result for $d_e/e$ lies between the two values
obtained from the naive computation, determining the heavy quark
masses at the different scales. All these results are illustrated in
Figure~\ref{fig:debottomcharm} of Section~\ref{sec:numerics}.

This paper is organised as follows: in Section~\ref{sec:eft}, we
introduce the effective theories used in the computation. In
Section~\ref{sec:rg}, we present the detailed results of our calculation,
namely, the initial conditions at the electroweak scale, the
calculation of the anomalous dimensions, and the threshold corrections
at the heavy-quark thresholds. We also show the final analytical
results in a compact form. In Section~\ref{sec:numerics}, we present the
numerical results of the calculation including updated bounds on the
CP-odd heavy-quark Yukawa couplings. We conclude in
Section~\ref{sec:conclusions}. Additional information is presented in two
appendices: In App.~\ref{app:unphys}, we collect the unphysical
operators used in the computation of the anomalous-dimension matrix
and in App.~\ref{app:ren_consts}, we show the results for
renormalisation constants entering the calculation.

\section{Effective Theories Below the Weak Scale\label{sec:eft}}

A precise determination of the electron EDM in the presence of CP-odd
Higgs couplings to the bottom and charm quarks requires an effective
field theory that allows to sum large QCD logarithms to all orders in
the strong coupling constant and includes the effect of the combined
$\alpha_s$ and $\alpha_e$ RG evolution.  Based on the ``full''
Lagrangian in Eq.~\eqref{eq:LagHqq}, we construct the effective
Lagrangian below the electroweak scale, $\muew$, by integrating out
the heavy degrees of freedom of the SM (the top quark, the weak gauge
bosons, and the Higgs).  EDMs are then induced by non-renormalisable
operators that are CP odd.  In the current work, we focus on the part
of the effective Lagrangian that is relevant for predicting the
electron EDM in the presence of CP-odd Yukawas couplings to the bottom
and charm quarks. In this case, the relevant effective Lagrangian
reads
\begin{equation} \label{eq:Leff}
  \Lag_{\text{eff},\q}= - \sqrt{2} \GF \, \Big(   C_1^{e\q} O_1^{e\q}
           + C_1^{\q e} O_1^{\q e}
           + C_2^{e\q} O_2^{e\q}
           + C_3^{e} O_3^{e}
    \Big) +\ldots \,,
\end{equation}
where the four linearly independent operators are
\begin{align}
  O_1^{e \q} &= (\bar e e) \, (\qb\, i \gamma_5 \q) \,,
  &O_1^{\q e} &= (\qb \q) \, (\bar e \, i \gamma_5 e) \,, \nonumber\\
  O_2^{e\q}  &= \frac{1}{2}\epsilon^{\mu\nu\rho\sigma} (\bar e \sigma_{\mu\nu} e) \, (\qb \, \sigma_{\rho\sigma} \q) \,,
  &O_3^e      &= \frac{Q_e}{2} \frac{m_\q}{e} \, (\bar e \sigma^{\mu \nu} e) \, \tilde{F}_{\mu \nu} \,,
\label{eq:dipoles}
\end{align}
and $C_1^{e\q}$, $C_1^{\q e}$, $C_2^{e\q}$, and $C_3^e$ are the
corresponding Wilson coefficients. Additional CP-odd operators that
are suppressed by additional factors of $m_\text{light}/m_\q$ (where
$m_\text{light}$ corresponds to a light quark mass) are denoted by the
ellipsis. We defined the electron dipole operator with a factor of the
running quark mass $m_\q \equiv m_\q(\mu)$, to avoid awkward ratios of
quark and lepton masses in the anomalous dimensions.  Throughout this
work the $\gamma_5$ matrix is defined by
\begin{equation}\label{eq:gamma5:def}
  \gamma_5 \equiv \frac{i}{4!} \epsilon_{\mu\nu\rho\sigma} \gamma^\mu
  \gamma^\nu \gamma^\rho \gamma^\sigma\,,
\end{equation}
where $\epsilon^{\mu\nu\rho\sigma}$ is the totally antisymmetric
Levi-Civita tensor in four space-time dimensions with
$\epsilon_{0123}=-\epsilon^{0123}=1$, and we use the notation
$\widetilde F^{\mu \nu} = \tfrac{1}{2} \epsilon^{\mu\nu\rho\sigma}
F_{\rho\sigma}$.  We treat $\gamma_5$ within the ``Larin'' scheme; for
the details and subtleties we refer to Ref.~\cite{Brod:2018pli}.  The
non-standard sign convention for $O_3^e$ is related to our definition
of the covariant derivative acting on fermion fields $f$,
\begin{equation}\label{eq:codev}
D_\mu \equiv \partial_\mu - i g_s T^a G_\mu^a + i e Q_f A_\mu \,.
\end{equation}

The basis of all flavour-diagonal, CP-odd operators is closed under
the QED and QCD RG flow, as both interactions conserve CP and
flavour. Below the electroweak scale, there is a tower of effective
theories relevant for predicting the electron EDM.  For the bottom
case, $\q=b$, we employ the effective Lagrangian in
Eq.~\eqref{eq:Leff} for the five-flavour theory, while for the charm
case, $\q = c$, we employ Eq.~\eqref{eq:Leff} for both the
five-flavour and the four-flavour theory (see discussion in Section
\ref{sec:thresh} for the threshold corrections on couplings and Wilson
coefficients at the bottom-quark scale).

The effective theory below the heavy-quark scale, $\mu_{\q}$, does not
contain four-fermion operators with heavy quarks, and we use the
modified effective Lagrangian
\begin{equation}
  \label{eq:LagTilde}
  \widetilde{\Lag}_{\text{eff},\q} = - \sqrt{2} \GF \widetilde{C}_3^e \widetilde{O}_3^e + \ldots \,,
\end{equation}
in which -- as opposed to Eq.~\eqref{eq:Leff} -- the dipole operator
is defined with the conventional factor $m_e$:
\begin{equation}\label{eq:O3e:below:mub}
  \widetilde{O}_3^e = \frac{Q_e}{2} \frac{m_e}{e} \, (\bar e \sigma^{\mu \nu} e) \, \tilde{F}_{\mu \nu} \,.
\end{equation}
Note that for $\q=b$ the Lagrangian in
Eq.~\eqref{eq:LagTilde} refers to the four-flavour Lagrangian, while
for $\q = c$ to the three-flavour one. This definition of
$\widetilde{\Lag}_{\text{eff},\q}$ implies that
(cf. Eq.~\eqref{eq:LeffN})
\begin{equation}\label{eq:de:match}
  \frac{d_e}{e} = - \sqrt{2} \GF \frac{m_e}{4\pi\alpha_e} \widetilde{C}_{3}^e \,.
\end{equation}
In the next section, we describe how the RG evolution within the
effective field theories relates $d_e$ to the parameters $\kappa_\q$
and $\phi_\q$ of the ``full'' theory in Eq.~\eqref{eq:LagHqq}.

\section{Renormalisation Group Evolution\label{sec:rg}}

Our goal is the summation of all leading and next-to-leading
logarithms via RG-improved perturbation theory. The calculation
proceeds in the following steps.

First, we integrate out the Higgs and weak gauge bosons together with
the top quark at the electroweak scale, $\muew \sim M_h$. This
matching calculation at $\muew$ induces the initial conditions for the
five-flavour Wilson coefficients appearing in Eq.~\eqref{eq:Leff}.  We
collect them in Section~\ref{sec:initialconditions}.  Subsequently, we
perform the RG evolution from $\muew$ down to the bottom-quark
threshold, $\mu_b \sim m_b(m_b)$.  The anomalous dimensions relevant
for the mixed QCD--QED RG evolution at NLL accuracy are computed here
for the first time, see Section~\ref{sec:anomalousdimensions}.  The
next step depends on whether $\q=b$ or $\q=c$.  For the bottom case,
$\q = b$, we match directly to the four-flavour version of the
Lagrangian in Eq.~\eqref{eq:LagTilde} to obtain the prediction of the
electron EDM.  The relevant threshold corrections at $\mu_b$ are
discussed in Section~\ref{sec:thresh}. For the charm case, $\q = c$,
we must instead match at $\mu_b$ to the four-flavour version of the
Lagrangian with four-fermion operators in Eq.~\eqref{eq:Leff} and
additionally perform the RG evolution in the four-flavour theory from
$\mu_b$ down to $\mu_c\sim m_c(m_c)$. The corresponding anomalous
dimensions are also given in
section~\ref{sec:anomalousdimensions}. Finally, we match to the
three-flavour version of the Lagrangian in Eq.~\eqref{eq:LagTilde} to
obtain the prediction of the electron EDM.

The calculations of the amplitudes relevant for computing the initial
conditions of Wilson coefficients and the hitherto unknown anomalous
dimensions have been performed with the package
\texttt{MaRTIn}~\cite{Brod:2024zaz} which is based on
\texttt{FORM}~\cite{Vermaseren:2000nd} and implements the two-loop
recursion algorithms presented in Refs.~\cite{Davydychev:1992mt,
  Bobeth:1999mk}. The amplitudes were generated using
\texttt{QGRAF}~\cite{Nogueira:1991ex}.

The RG evolution between the different matching scales is
significantly more involved than in applications in which the
electromagnetic coupling, $\alpha_e$, can be neglected. The reason is
that the leading contribution to the electron EDM contains a term with
two powers of the large logarithm, i.e, $\log^2x_\q$, see
Eq.~\eqref{eq:dsw}.  Within the effective theory, this term is
obtained by a LL QED calculation, truncated at order $\alpha_e^2$.
However, the numerically relevant corrections to this result are not
further electromagnetic $\alpha_e^n\log^{n}$ corrections, but large,
logarithmically enhanced QCD corrections which must be summed to all
orders to obtain an accurate prediction.  Therefore, to properly
account for the numerically relevant corrections we must consistently
solve the mixed QCD--QED RG equations.

This can be achieved consistently using the general formalism
developed in Ref.~\cite{Huber:2005ig}.  The main idea is that, since
$\alpha_s\log x_\q$ is ${\cal O}(1)$, such products of QCD coupling
times large logarithms must be resummed for an accurate prediction. By
contrast, the product $\alpha_e\log x_\q$ is small and does not
require resummation. The large logarithms appearing in
$\alpha_e\log x_\q$ are expressed in terms of the resummed
$\alpha_s\log x_\q$, i.e.
$\alpha_e\log x_\q = (\alpha_e/\alpha_s) \times\alpha_s\log x_\q$.  As
a consequence, the conventional expansion in terms of $\alpha_s$ and
$\alpha_e$ is replaced by an expansion in $\alpha_s$ and
$\kappa \equiv \alpha_e /\alpha_s$.  Therefore, in this setup the
Wilson coefficients at some scale $\mu$ have the expansion
\begin{equation}
  \label{eq:Caskappa}
  C_i(\mu) = \sum_{n, m = 0}^\infty \tilde{\alpha}_s(\mu)^n\kappa(\mu)^m\,C^{(nm)}_i(\mu)\,,
\end{equation}
with $\tilde{\alpha}_{s} \equiv \alpha_{s}/4\pi$.  By implementing the
mixed RG evolution we will compute the $C^{(nm)}_i(\mu)$ coefficients
relevant for the electron EDM; for details we refer to
Ref.~\cite{Huber:2005ig}.  The analytical results are presented in
Section~\ref{sec:anres}.

\subsection{Initial Conditions at the Weak Scale\label{sec:initialconditions}}

\begin{figure}[t]
  \centering
  \includegraphics{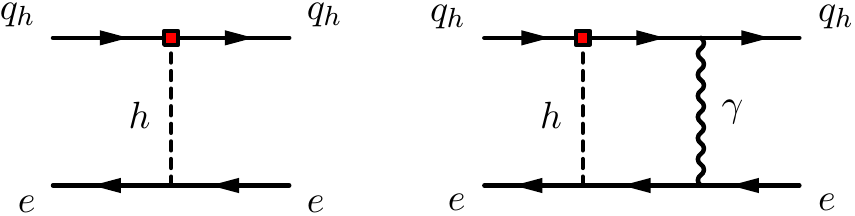}
	\caption{Examples of leading order (left) and next-to-leading order (right)
	Feynman diagrams that contribute to the initial conditions of the four-fermion
	sector in the EFT. CP-odd heavy-quark Yukawa couplings are
	indicated by red squares.
	\label{fig:initialconditions}}
\end{figure}

We augment the SM by flavour-conserving, CP-odd Higgs Yukawa couplings
to the heavy quarks $\q =b,c$, as parameterised in
Eq.~\eqref{eq:LagHqq}.  At a scale $\muew\approx M_h$ we integrate out
the heavy degrees of freedom of the SM and match to the effective
five-flavour theory relevant for the electron EDM in
Eq.~\eqref{eq:Leff}. The initial conditions for the Wilson
coefficients relevant for our calculation are obtained by evaluating
the tree-level and one-loop Feynman diagrams such as those shown in Figure~\ref{fig:initialconditions}; we find
\begin{align}
C_1^{e\q}(\muew) &= - \kappa_\q \sin\phi_\q \frac{m_e m_{\q}}{M_h^2}
                  + {\cal O} (\alpha_s^2, \alpha_s \alpha_e, \alpha_e^2) \,,\label{eq:init:C1eb}\\
C_1^{\q e}(\muew) &= 0 
                  + {\cal O} (\alpha_s^2, \alpha_s \alpha_e, \alpha_e^2) \,,\label{eq:init:C1be}\\
\begin{split}
C_2^{e\q}(\muew) &=  \frac{\alpha_e}{4\pi}
         \left(\frac{3}{2} + \log\frac{\muew^2}{M_h^2} \right)\frac{m_e m_{\q}}{M_h^2}
	Q_e Q_\q \kappa_\q \sin\phi_\q
    + {\cal O} (\alpha_s^2, \alpha_s \alpha_e, \alpha_e^2)\,,
\end{split}\label{eq:init:C2eb}\\
C_3^{e}(\muew)    &= 0 + {\cal O} (\alpha_e^2)\,.
\end{align}
All parameters appearing above correspond to parameters in the
five-flavour theory evaluated at the scale $\muew$, i.e.,
$m_\q = m_{\q,\text{5fl}}(\muew)$ and
$\alpha_e = \alpha_{e,\text{5fl}}(\muew)$. We treat the heavy quarks
and the electron as massless at the electroweak matching scale;
therefore, no powers of $m_e/M_h$ or $m_\q/M_h$ appear in the
results. The explicit factors of the electron and heavy-quark masses
arise by expressing the Yukawa couplings in Eq.~\eqref{eq:LagHqq} in
terms of $m_e$ and $m_\q$. We have included all terms up to
corrections of quadratic order in the strong and electromagnetic
coupling constants, as only these are required for our analysis. The
${\mathcal O}(\alpha_e)$ coefficient, Eq.~\eqref{eq:init:C2eb}, is
well-defined only after specifying the basis of evanescent operators
which can be found in App.~\ref{app:unphys}.

In Section~\ref{sec:anres}, we will use the framework of
Ref.~\cite{Huber:2005ig} to solve the mixed QCD--QED RG as an
expansion in $\tilde{\alpha}_s$ and $\kappa$. Based on
Eqs.~\eqref{eq:init:C1eb}-\eqref{eq:init:C2eb} we find the
contributing expansion coefficients in Eq.~\eqref{eq:Caskappa}
\begin{equation}
	\begin{split}
      C_1^{e \q,(00)}(\muew) &= - \kappa_\q \sin\phi_\q \frac{m_e m_{\q}}{M_h^2}\,, \\
		C_2^{e \q,(11)}(\muew) &= \left(\frac{3}{2} + \log\frac{\muew^2}{M_h^2} \right)\frac{m_e m_{\q}}{M_h^2}
        Q_e Q_\q \kappa_\q \sin\phi_\q \,.
	\end{split}
\end{equation}

\subsection{Anomalous Dimensions\label{sec:anomalousdimensions}}

To solve the RG in the effective theories below the electroweak scale
we need to include the running of coupling constants, mass anomalous
dimensions, and the mixing of operators. In this section, we collect
the results that enter the analysis at NLL accuracy.

The running of the Wilson coefficients from the electroweak matching
scale down to the relevant quark scale is governed by the RG equation
\begin{equation}
  \frac{dC_i}{d\log\mu} = C_j \gamma_{ji}\,,
\end{equation}
where $\gamma_{ji}$ are the components of the anomalous dimension
matrix (ADM). We choose the ordering of Wilson coefficients as
\begin{equation}
  \vec{C} = (C_1^{e\q},~C_1^{\q e},~C_2^{e\q},~C_3^e)^T\,.
\end{equation}
The ADM admits an expansion in powers of\footnote{ Note that for the
  ADM we do not expand in terms of $\tilde{\alpha}_s$ and $\kappa$ as
  for Wilson coefficients.} $\tilde{\alpha}_s\equiv \alpha_s/4\pi$ and
$\tilde{\alpha}_e\equiv \alpha_e/4\pi$,
\begin{equation}
  \gamma = \sum_{\substack{n,m=0\\n+m\geq 1}} \gamma^{(nm)}
  \tilde{\alpha}_s(\mu)^n\tilde{\alpha}_e(\mu)^m \,,
\end{equation}
with $\tilde{\alpha}_s$, $\tilde{\alpha}_e$, and $\gamma^{(nm)}$
depending on the number of active fermion flavours in the effective
theory.  Using this expansion, the ADM can be organised by loop order.
In general, it depends on the number of active fermion flavours and on
the flavour of the heavy quark $\q$. Below we present our results
entering the five- and four-flavour RG evolution required for the case
$\q=b$ and $\q=c$.

By explicit calculation, we find at one loop
\begin{align}
\label{eq:adm01}
\gamma^{(01)}&=
\begin{pmatrix}
-6(Q_e^2+Q_\q^2)&0&-2Q_e Q_\q&0\\[0.5em]
0&-6(Q_e^2+Q_\q^2)&-2Q_e Q_\q&0\\[0.5em]
-48Q_\q Q_e&-48Q_\q Q_e&2(Q_e^2+Q_\q^2)& - 48 \frac{Q_\q}{Q_e}\\[0.5em]
0&0&0&10Q_e^2+6Q_\q^2-2\beta_e^{(0)}
\end{pmatrix}\,,\\[10pt]
\gamma^{(10)}&=
\begin{pmatrix}
-6C_F&0&0&0\\[0.5em]
0&-6C_F&0&0\\[0.5em]
0&0&2C_F&0\\[0.5em]
0&0&0&\gamma_{\q,s}^{(0)}
\end{pmatrix}\,,
\label{eq:adm10}
\end{align}
with
$\beta_e^{(0)} = -\frac{4}{3} \Big(N_c n_d Q_b^2+N_c n_u Q_c^2+n_\ell
Q_e^2\Big)$ and $\gamma_{\q,s}^{(0)}=6 C_F$. Moreover, $N_c = 3$ is
the number of quark colors, $n_u$ is the number of active up-type
quarks, $n_d$ is the number of active down-type quarks, $n_\ell$ is
the number of active charged leptons and
$C_F \equiv (N_c^2 - 1)/2N_c = 4/3$.  In both the bottom and charm
cases, we have $n_u = 2$ and $n_\ell = 3$.

\begin{figure}[t]
  \centering
  \includegraphics{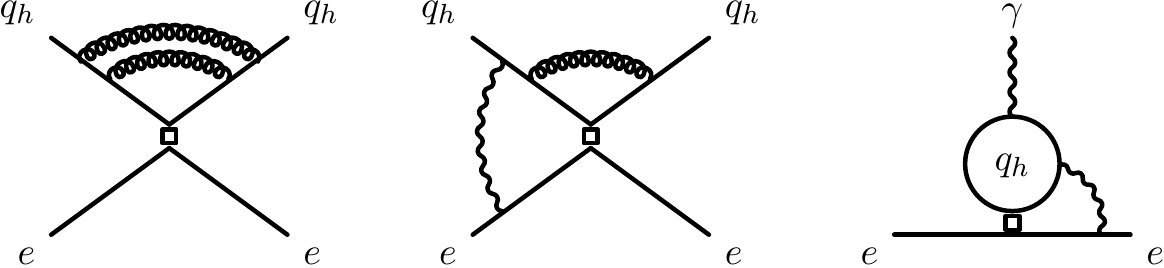}
	\caption{Examples of two-loop QCD (left), mixed QED--QCD (middle), and
	two-loop QED (right) Feynman diagrams entering the computation of the
	anomalous dimension matrix. Effective operator insertions are represented
	by the unshaded boxes. \label{fig:adm}}
\end{figure}

At two-loop, the pure QCD ADM is the same for the bottom- and
charm-quark case and depends on the number of active quark flavours,
\begin{equation}\label{eq:adm:qcd:2}
\gamma^{(20)}
=
\begin{pmatrix}
\frac{724}{9} - \frac{88n_d}{9}&0&0&0\\[0.5em]
0&-\frac{1132}{9} + \frac{40n_d}{9}&0&0\\[0.5em]
0&0&\frac{1964}{27}-\frac{104 n_d}{27}&0\\[0.5em]
0&0&0&\frac{1132}{9} - \frac{40n_d}{9}
\end{pmatrix} \,.
\end{equation}
For the bottom case, we only need the five-flavour theory, i.e., we
always fix $n_d = 3$.  For the charm case, we must solve the RG in
both five- and four-flavour theory in which case $n_d = 3$ and
$n_d = 2$, respectively.

For the two-loop mixed QCD--QED and the pure two-loop QED ADMs,
explicit factors of the quark charges result in different results for
the bottom and charm cases. For notational simplicity, we quote the
ADMs for the two cases separately after having substituted the
electric charges.  For the bottom, we find, after fixing $n_d = 3$
\begin{equation}\label{eq:adm:2}
  \gamma^{(11)}_{[\q=b]}
=
\begin{pmatrix}
-\frac{8}{9}&0&-\frac{16}{9}&0\\[0.5em]
0&-\frac{8}{9}&\frac{16}{3}&0\\[0.5em]
128&-\frac{128}{3}&-\frac{152}{27}&-\frac{448}{3}\\[0.5em]
0&0&0&40
\end{pmatrix} \,, \quad
\gamma^{(02)}_{[\q=b]}
=
\begin{pmatrix}
\frac{2420}{81}&-\frac{14}{3}&\frac{392}{81}&-8\\[0.5em]
-\frac{14}{3}&-\frac{7820}{81}&\frac{8}{81}&-\frac{8}{3}\\[0.5em]
\frac{64}{27}&\frac{3136}{27}&-\frac{17234}{243}&\frac{464}{9}\\[0.5em]
0&0&0&-\frac{8704}{81}
\end{pmatrix}\,.
\end{equation}
For the charm, we find, keeping the $n_d$-dependence explicit
\begin{align}
\gamma^{(11)}_{[\q=c]} &=
  \begin{pmatrix}
  -\frac{32}{9}&0&\frac{32}{9}&0\\[0.5em]
  0&-\frac{32}{9}&-\frac{32}{3}&0\\[0.5em]
  -256&\frac{256}{3}&-\frac{608}{27}&\frac{896}{3}\\[0.5em]
  0&0&0&32 + \frac{32 n_d}{9}
  \end{pmatrix} \,,\\[10pt]
  \gamma^{(02)}_{[\q=c]}&=
\begin{pmatrix}
 -\frac{439}{81}+\frac{4n_d}{81}&-\frac{56}{3}&-\frac{688}{81}-\frac{8n_d}{81}&-32\\[0.5em]
 -\frac{56}{3}&-\frac{5879}{81}-\frac{316n_d}{81}&-\frac{208}{81}-\frac{8n_d}{81}&-\frac{32}{3}\\[0.5em]
-\frac{1664}{27}-\frac{64n_d}{27}&-\frac{5504}{27}-\frac{64n_d}{27}&
-\frac{25661}{243}-\frac{676n_d}{243}&-\frac{256}{9}\\[0.5em]
0&0&0&-\frac{8647}{81}-\frac{404 n_d}{81}
\end{pmatrix}\,.
\end{align}
Some representative Feynman diagrams used to compute the two-loop ADMs
are shown in Figure~\ref{fig:adm}. The two-loop ADM are well-defined
only after specifying the basis of evanescent operators; this is done
in App.~\ref{app:unphys}.

All two-loop results in this section are new, to the best of our
knowledge. The one-loop RG evolution has also been calculated in
Ref.~\cite{Jenkins:2017dyc}.

\subsection{Threshold Corrections\label{sec:thresh}}

When performing the running of the Wilson coefficients, we must also
integrate out the heavy quarks at the respective scales. This leads to
threshold corrections to the gauge couplings and quark masses, as well
as to the Wilson coefficients. 
% The threshold corrections must be
% consistently taken into account to cancel the spurious dependence of
% the final result on the matching scales.
%
Below we describe the origin of the threshold corrections for each
case, and collect the corresponding results. Due to the mixed RG
evolution, the matching of the effective theories must also be
performed as an expansion in $\tilde{\alpha}_s$ and $\kappa$,
cf. Eq.~\eqref{eq:Caskappa}. We stress that the results presented
below are only applicable for our specific case in which the only
non-vanishing initial conditions are the ones in
section~\ref{sec:initialconditions}.  Other UV completions, with more
contributions to the initial conditions, can receive additional
threshold corrections (for instance, if $C^e_3(\muew)\neq 0$ at
one-loop).

\subsubsection*{The Bottom-Quark Case}

In the case of an anomalous bottom-quark Yukawa coupling, $\q=b$, the
only relevant threshold below the weak scale is at $\mu_b$ in which we
match the five-flavour version of the Lagrangian in
Eq.~\eqref{eq:Leff} to the four-flavour version of the Lagrangian in
Eq.~\eqref{eq:LagTilde}.  There are three effects that induce a
non-trivial correction to the matching onto $\widetilde{C}_3^e$: the
threshold correction for $\alpha_s$ when matching from the
five-flavour onto the four-flavour theories (the corresponding one for
$\alpha_e$ does not contribute because in our case
$C_{3,\text{5fl}}^{e,(01)}(\mu_b)=0$); the different normalization of
the dipole operators $O^e_3$ and $\widetilde{O}^e_3$ in the two
theories, which leads to a factor of $m_b/m_e$; and a threshold
correction from one-loop insertions of $O^{be}_2$ in the five-flavour
theory. Details can be found in the analogous discussion in
Ref.~\cite{Brod:2018pli}.

Taking all of these effects into account, we find for the electron
dipole operator in the four-flavour theory at $\mu_b$
\begin{align}
  \widetilde{C}_{3}^{e}(\mu_b) = &
  \kkappa{4}^2 \frac{m_b}{m_e} C_{3,\text{5fl}}^{e,(02)}(\mu_b)\nonumber\\
  +&{\alphast{4}} \kkappa{4}^2 \frac{m_b}{m_e}
  \biggl(C_{3,\text{5fl}}^{e,(12)}(\mu_b)
    + 24 \frac{Q_b}{Q_e}\log\frac{\mu_b^2}{m_b^2}C_{2,\text{5fl}}^{be,(01)}(\mu_b)
     \nonumber\\
   &\qquad\qquad\qquad\qquad- \frac{1}{2}\gamma_{b,s}^{(0)} \log\frac{\mu_b^2}{m_b^2}C_{3,\text{5fl}}^{e,(02)}(\mu_b)
     - 2 \delta\alpha_s          \log\frac{\mu_b^2}{m_b^2}C_{3,\text{5fl}}^{e,(02)}(\mu_b)\biggr)\\
  =&
  \kkappa{4}^2 \frac{m_b}{m_e} C_{3,\text{5fl}}^{e,(02)}(\mu_b)\nonumber\\
        +&{\alphast{4}} \kkappa{4}^2 \frac{m_b}{m_e}
  \biggl(C_{3,\text{5fl}}^{e,(12)}(\mu_b)
        + 8\log\frac{\mu_b^2}{m_b^2}C_{2,\text{5fl}}^{be,(01)}(\mu_b)
        - \frac{16}{3} \log\frac{\mu_b^2}{m_b^2}C_{3,\text{5fl}}^{e,(02)}(\mu_b)\biggr)\,,
  \label{eq:thresholdBmub}
\end{align}
with $\gamma_{b,s}^{0} = 6 C_F$ the leading-order QCD mass anomalous
dimension, the $\alpha_s$ threshold correction $\delta\alpha_s =2/3$,
and $m_b = m_b(m_b)$ in the five-flavour theory in the
$\overline{\text{MS}}$ scheme. We indicate by explicit subscripts
``5fl'' and ``4fl'' in which effective theory the various quantities
are defined. The couplings in Eq.~\eqref{eq:thresholdBmub} are
evaluated at $\mu_b$, i.e., $\alphast{4}(\mu_b)$ and
$\kkappa{4}(\mu_b)$.

\subsubsection*{The Charm-Quark Case}

In the case of an anomalous charm-quark Yukawa coupling, $\q=b$, there
are threshold corrections both at the bottom- and the charm-quark
thresholds, $\mu_b$ and $\mu_c$, respectively.

At $\mu_b$, the effective Lagrangian in the five- and four-flavour
theories are the same (see Eq.~\eqref{eq:Leff}), and at NLL accuracy
the only relevant effect is the decoupling of $\alpha_s$.  The only
non-trivial matching condition at the bottom-quark scale then reads
for the charm-quark case, $\q=c$, 
\begin{equation}
        C_{3,\text{4fl}}^{e}(\mu_b) = \kkappa{4}^2 C_{3,\text{5fl}}^{e,(02)}(\mu_b) + 
                                      \alphast{4}\kkappa{4}^2\biggl(
                                      C_{3,\text{5fl}}^{e,(12)}(\mu_b)
                                    - 2\delta \alpha_s \log\frac{\mu_b^2}{m_b^2}C_{3,\text{5fl}}^{e,(02)}(\mu_b)\biggr)\,.
\end{equation}
Other operators beyond $O_3^e$ also receive threshold
corrections. However, none of these terms enter the final
three-flavour value of $\widetilde{C}_{3}^{e}(\mu_c)$ at NLL order, so
we do not list them here.

At the charm-quark threshold, $\mu_c$, the threshold correction from
matching the four-flavour onto the three-flavour theory with the
single operator in Eq.~\eqref{eq:LagTilde} is analogous to the bottom
case. Accordingly, we find for $\q=c$
\begin{align}
	\widetilde{C}_{3}^{e}(\mu_c) = &
  \kkappa{3}^2 \frac{m_c}{m_e} C_{3,\text{4fl}}^{e,(02)}(\mu_c)\nonumber\\
  +&{\alphast{3}} \kkappa{3}^2 \frac{m_c}{m_e}
  \biggl(C_{3,\text{4fl}}^{e,(12)}(\mu_c)
    + 24 \frac{Q_c}{Q_e}\log\frac{\mu_c^2}{m_c^2}C_{2,\text{4fl}}^{ce,(01)}(\mu_c)
     \nonumber\\
     &\qquad\qquad\qquad\qquad- \frac{1}{2}\gamma_{c,s}^{(0)} \log\frac{\mu_c^2}{m_c^2}C_{3,\text{4fl}}^{e,(02)}(\mu_c)
     - 2 \delta\alpha_s\log\frac{\mu_c^2}{m_c^2}C_{3,\text{4fl}}^{e,(02)}(\mu_c)\biggr)\\
  =&
  \kkappa{3}^2 \frac{m_c}{m_e} C_{3,\text{4fl}}^{e,(02)}(\mu_c)\nonumber\\
 +&{\alphast{3}} \kkappa{3}^2 \frac{m_c}{m_e}
  \biggl(C_{3,\text{4fl}}^{e,(12)}(\mu_c) 
        - 16\log\frac{\mu_c^2}{m_c^2}C_{2,\text{4fl}}^{ce,(01)}(\mu_c)
      - \frac{16}{3} \log\frac{\mu_c^2}{m_c^2}C_{3,\text{4fl}}^{e,(02)}(\mu_c)\biggr)\,,
\end{align}
where $\gamma_{c,s}^{0} = 6 C_F$, $\delta\alpha_s =2/3$, and
$m_c = m_c(m_c)$ in the four-flavour theory in the
$\overline{\text{MS}}$ scheme.

\subsection{Analytic Solution of the RG \label{sec:anres}}

In this section, we show the final result for the electron EDM Wilson
coefficients after implementing the mixed RG evolution (see discussion
above Eq.~\eqref{eq:Caskappa}) using the ADMs from
section~\ref{sec:anomalousdimensions}, and including the threshold
corrections from section~\ref{sec:thresh}.  We obtain obtain the exact
analytical result for the contribution to the electron dipole moment
up to $\mathcal{O}(\alpha_s\kappa^2)$ including the resummation of QCD
logarithms:
\begin{equation}
	\frac{d_e}{e}
		=  - \sqrt{2}G_{\rm F} \frac{m_e}{4\pi \alpha_e}
		  \left[ 
                              \kappa^2 \widetilde{C}_{3}^{e,(02)}(\mu_\q)
            +\tilde{\alpha}_s \kappa^2 \widetilde{C}_{3}^{e,(12)}(\mu_\q)
          +{ \cal O}\big(\tilde{\alpha}_s^2\kappa^2,\kappa^3\big)\right]\,,
          \label{eq:dresum}
\end{equation}
with $\alpha_e$, $\tilde{\alpha}_s$, and
$\kappa\equiv \alpha_e/\alpha_s$ evaluated at the scale $\mu_\q$ in
the four- and three-flavour theory for the bottom- and charm-quark
case, respectively.  The coefficients
$\widetilde{C}_{3}^{e,(mn)}(\mu_\q)$ are functions of the initial
conditions at $\muew$ and of ratios of the values of $\alpha_s$ at
different scales.  For brevity we introduce corresponding compact
notations
\begin{align*}
  \bar{C}_{iX}^{(mn)} \equiv {C}_{i,\text{5fl}}^{X,(mn)}(\muew)\,,
  \qquad
  \eta_5 \equiv \frac{\alphas{5}(\muew)}{\alphas{5}(\mu_b)}\,,
  \qquad
  \eta_4 \equiv \frac{\alphas{4}(\mu_b)}{\alphas{4}(\mu_c)}\,.
\end{align*}
In both the bottom and charm-quark cases, the operator
$\widetilde{O}_e^3$ in Eq.~\eqref{eq:LagTilde} below the respective
heavy-quark scale is modified only by QED running effects which are
negligibly small.

\subsubsection*{Bottom-Quark Result}
In the bottom-quark case, $\q=b$, we find for the coefficients in Eq.~\eqref{eq:dresum}
\begin{align}
  \widetilde{C}_{3}^{e,(02)}(\mu_b)&=
    \frac{m_b}{m_e}
    %(\bar C_{1be}^{(00)} + \bar C_{1eb}^{(00)})
    \bar C_{1eb}^{(00)}
    \biggl(\frac{12}{77} \eta_5^{-\frac{12}{23}}-\frac{8}{35} \eta_5^{-\frac{19}{23}}+\frac{4}{55} \eta_5^{-\frac{34}{23}}\biggr)\,,\\[1em]
    \widetilde{C}_{3}^{e,(12)}(\mu_b)&=
    \frac{m_b}{m_e}
    %(\bar C_{1be}^{(00)} + \bar C_{1eb}^{(00)})
    \bar C_{1eb}^{(00)}
    \left(\frac{248}{105} \eta_5^{-\frac{19}{23}}-\frac{152}{77} \eta_5^{-\frac{12}{23}}-\frac{64}{165} \eta_5^{-\frac{34}{23}}\right)
    \log\frac{\mu_b^2}{m_b^2}\nonumber\\
    %&+
    %\frac{m_b}{m_e}
    %\bar C_{1be}^{(00)} \biggl(+\frac{13138}{23805} \eta_5^\frac{4}{23}-\frac{4264}{5819} \eta_5^\frac{11}{23}+\frac{5671552}{261855} \eta_5^{-\frac{11}{23}}\nonumber\\
    %&\qquad\qquad\qquad~~\quad-\frac{3022034}{122199} \eta_5^{-\frac{12}{23}}+\frac{199636}{55545} \eta_5^{-\frac{19}{23}}-\frac{29848}{87285} \eta_5^{-\frac{34}{23}}\biggr)\nonumber\\
    &+
    \frac{m_b}{m_e}
    \bar C_{1eb}^{(00)} \biggl(\frac{37864}{40733} \eta_5^\frac{11}{23}-\frac{314306}{166635} \eta_5^\frac{4}{23}+\frac{3779848}{261855} \eta_5^{-\frac{11}{23}}\nonumber\\
    &\qquad\qquad~~\quad-\frac{2044442}{122199} \eta_5^{-\frac{12}{23}}+\frac{199636}{55545} \eta_5^{-\frac{19}{23}}-\frac{29848}{87285} \eta_5^{-\frac{34}{23}}\biggr)\nonumber\\
    &+
    \frac{m_b}{m_e}
    \bar C_{2eb}^{(11)} \biggl(\frac{8}{5} \eta_5^{-\frac{11}{23}}-\frac{8}{5} \eta_5^\frac{4}{23}\biggr)\,,
  \label{eq:masterbottom}
\end{align}
where $m_b = m_b(m_b)$ in the five-flavour theory and in the
$\overline{\textrm{MS}}$ scheme.

\subsubsection*{Charm-Quark Result}
In the charm-quark case, $\q=c$, we find for the coefficients in Eq.~\eqref{eq:dresum}
\begin{align}
\widetilde{C}_{3}^{e,(02)}(\mu_c)&=
\frac{m_c}{m_e}\bar C_{1ec}^{(00)} \biggl(
 \frac{16}{39} \eta_4^{-\frac{12}{25}} \eta_5^{-\frac{12}{23}}
+\frac{64}{357} \eta_4^{-\frac{21}{25}} \eta_5^{-\frac{12}{23}}
+\frac{576}{17017} \eta_4^{-\frac{38}{25}} \eta_5^{-\frac{12}{23}}\nonumber\\
&\qquad\qquad~ ~
-\frac{96}{119} \eta_4^{-\frac{21}{25}} \eta_5^{-\frac{19}{23}}
-\frac{64}{595} \eta_4^{-\frac{38}{25}} \eta_5^{-\frac{19}{23}}
+\frac{16}{55} \eta_4^{-\frac{38}{25}} \eta_5^{-\frac{34}{23}}\biggr)\,,\\[1em]
\widetilde{C}_{3}^{e,(12)}(\mu_c)&=
\frac{m_c}{m_e}\bar C_{1ec}^{(00)} \log\frac{\mu_b^2}{m_b^2} \biggl(
 \frac{64}{119} \eta_4^{\frac{4}{25}} \eta_5^{-\frac{19}{23}}
-\frac{64}{119} \eta_4^{\frac{4}{25}} \eta_5^{-\frac{12}{23}}
-\frac{384}{1309} \eta_4^{-\frac{13}{25}} \eta_5^{-\frac{12}{23}}
\nonumber\\&\qquad\qquad\qquad\quad~~~
+\frac{1216}{1785} \eta_4^{-\frac{13}{25}} \eta_5^{-\frac{19}{23}}
-\frac{64}{165} \eta_4^{-\frac{13}{25}} \eta_5^{-\frac{34}{23}}
\biggr)\nonumber\\
&+\frac{m_c}{m_e}\bar C_{1ec}^{(00)} \log \frac{\mu_c^2}{m_c^2}\biggl(
 \frac{1056}{119} \eta_4^{-\frac{21}{25}} \eta_5^{-\frac{19}{23}}
-\frac{224}{39} \eta_4^{-\frac{12}{25}} \eta_5^{-\frac{12}{23}}
-\frac{704}{357} \eta_4^{-\frac{21}{25}} \eta_5^{-\frac{12}{23}}
\nonumber\\&\qquad\qquad\qquad\quad~~~
-\frac{3072}{17017} \eta_4^{-\frac{38}{25}} \eta_5^{-\frac{12}{23}}
+\frac{1024}{1785} \eta_4^{-\frac{38}{25}} \eta_5^{-\frac{19}{23}}
-\frac{256}{165} \eta_4^{-\frac{38}{25}} \eta_5^{-\frac{34}{23}}
\biggr)\nonumber\\
&+\frac{m_c}{m_e}\bar C_{1ec}^{(00)} \biggl(
 \frac{605824}{566559} \eta_4^{\frac{4}{25}} \eta_5^{\frac{11}{23}}
-\frac{1257224}{188853} \eta_4^{\frac{4}{25}} \eta_5^{\frac{4}{23}}
+\frac{151456}{61893} \eta_4^{\frac{13}{25}} \eta_5^{\frac{11}{23}}
\nonumber\\&\qquad\qquad~~
-\frac{21158428}{16861875} \eta_4^{\frac{4}{25}} \eta_5^{-\frac{12}{23}}
+\frac{3414272}{12894375} \eta_4^{\frac{13}{25}} \eta_5^{-\frac{12}{23}}
+\frac{891392}{5620625} \eta_4^{\frac{4}{25}} \eta_5^{-\frac{19}{23}}
\nonumber\\&\qquad\qquad~~
-\frac{2514448}{2832795} \eta_5^{\frac{4}{23}} \eta_4^{-\frac{13}{25}}
+\frac{1817472}{9001993} \eta_5^{\frac{11}{23}} \eta_4^{-\frac{13}{25}}
+\frac{15119392}{261855} \eta_4^{-\frac{13}{25}} \eta_5^{-\frac{11}{23}}
\nonumber\\&\qquad\qquad~~
+\frac{8657588}{219375} \eta_4^{-\frac{12}{25}} \eta_5^{-\frac{12}{23}}
-\frac{578877793792}{5626245625} \eta_4^{-\frac{13}{25}} \eta_5^{-\frac{12}{23}}
-\frac{4759984}{2008125} \eta_4^{-\frac{21}{25}} \eta_5^{-\frac{12}{23}}
\nonumber\\&\qquad\qquad~~
-\frac{1460352}{10635625} \eta_4^{-\frac{38}{25}} \eta_5^{-\frac{12}{23}}
+\frac{612782088}{196721875} \eta_4^{-\frac{13}{25}} \eta_5^{-\frac{19}{23}}
+\frac{2379992}{223125} \eta_4^{-\frac{21}{25}} \eta_5^{-\frac{19}{23}}
\nonumber\\&\qquad\qquad~~
+\frac{486784}{1115625} \eta_4^{-\frac{38}{25}} \eta_5^{-\frac{19}{23}}
-\frac{3414272}{18184375} \eta_4^{-\frac{13}{25}} \eta_5^{-\frac{34}{23}}
-\frac{121696}{103125} \eta_4^{-\frac{38}{25}} \eta_5^{-\frac{34}{23}}
\biggr)\nonumber\\
&+\frac{m_c}{m_e}\bar C_{2ec}^{(11)} \biggl(
 \frac{48}{17} \eta_4^{\frac{4}{25}} \eta_5^{\frac{4}{23}}
+\frac{32}{85} \eta_5^{\frac{4}{23}} \eta_4^{-\frac{13}{25}}
-\frac{16}{5} \eta_4^{-\frac{13}{25}} \eta_5^{-\frac{11}{23}}
\biggr)
\end{align}
where $m_c = m_c(m_c)$ in the four-flavour theory and in the $\overline{\textrm{MS}}$ scheme.

\section{Numerical Results\label{sec:numerics}}

In this section, we present the numerical results based on the
analytic expressions of the previous section and obtain constraints on
CP-odd Higgs Yukawas to the bottom and charm quarks from the electron
EDM.  Combining the initial conditions in
section~\ref{sec:initialconditions} with the RG solution in
section~\ref{sec:anres} leads to the electron EDM prediction, cf.,
Eq.~\eqref{eq:de:match},
  \begin{multline}
	\frac{d_e}{e}
    =  \frac{\sqrt{2}G_{\rm F}}{4\pi \alpha_e}m_e\times
    \left\{
      \begin{array}{l}
          \dfrac{m_b(m_b) m_b(\muew)}{M_h^2}
          \kappa_b\sin\phi_b
		  \Big[ 
            \kappa^2 F^{\text{LL}}_b(\eta_5)\\
          \hspace*{8em}\quad\,          +\tilde{\alpha}_s \kappa^2 F^{\text{NLL}}_b(\eta_5;\log\frac{\mu_b}{m_b},\log\frac{\muew}{M_h})\Big]\\[1em]
            \dfrac{m_c(m_c) m_c(\muew)}{M_h^2}
          \kappa_c\sin\phi_c
		  \Big[ 
            \kappa^2 F^{\text{LL}}_c(\eta_5,\eta_4)\\
          \hspace*{8em}\quad\,+\tilde{\alpha}_s \kappa^2 F^{\text{NLL}}_c(\eta_4,\eta_5;\log\frac{\mu_c}{m_c},\log\frac{\mu_b}{m_b},\log\frac{\muew}{M_h})\Big]
      \end{array}
    \right\}\\+ {\cal O}\big(\tilde{\alpha}_s^2\kappa^2,\kappa^3\big)\,.
  \end{multline}
where the heavy quark masses at the electroweak scale are given by
their NLL running relations
\begin{equation}
	\begin{split}
		&\frac{m_b(\muew)}{m_b(m_b)} = \eta_5^{\frac{12}{23}}\Bigg[1 
		+ \frac{\alpha_{s,5\text{fl}}(\mu_b)}{4\pi}\Bigg(\frac{7462}{1587}\big(\eta_5 - 1\big)
		- 4\log\frac{\mu_b^2}{m_b^2}\Bigg)\Bigg]\,,
			\\[5pt]
		&\frac{m_c(\muew)}{m_c(m_c)} = \eta_4^{\frac{12}{25}}\eta_5^{\frac{12}{23}}\Bigg[1
		+ \frac{\alpha_{s,4\text{fl}}(\mu_c)}{4\pi}\Bigg(\frac{7462}{1587}\eta_4\eta_5
		- \frac{213392}{330625}\eta_4 - \frac{7606}{1875} - 4\log\frac{\mu_c^2}{m_c^2}\Bigg)\Bigg]\,.
	\end{split}
\end{equation}
This result demonstrates explicitly how our RG-improved calculation
removes the ambiguity of the fixed-order computation by resumming the
large $\alpha_s$ logarithms and thus defining the scale at which the
masses are evaluated: in contrast to the fixed-order result in
Eq.~\eqref{eq:dsw}, $d_e$ is here proportional to
$m_\q(m_\q) m_\q(\muew) /M_h^2$ and the functions
$F_{\q}^{\text{(N)LL}}$, which contain the $\alpha_s$ resummation and
do not depend on the large logarithms $\log m_\q/M_h$. Expanding the
LL result of Eq.~\eqref{eq:dresum} in $\alpha_s(\muew)$ we recover
exactly the $\log^2 x_q$ term in Eq.~\eqref{eq:dsw}. On the other
hand, we cannot reproduce the $\pi^2/3$ term in Eq.~\eqref{eq:dsw}
since it must arise from the higher-order terms in
Eq.~\eqref{eq:dresum} of order $\kappa^2 \tilde \alpha_s^2$, which are
counted as NNLL in the QCD resummation.

\begin{table}[t]
  \begin{center}
    \caption{Input parameters used in evaluating the low-scale Wilson 
      coefficient $\widetilde{C}_3^e(\mu_\q)$  and equivalently $d_e/e$.
    All values are taken from Ref.~\cite{Zyla:2020zbs}; running parameters
	  are evaluated in the $\overline{\text{MS}}$ scheme.
          \label{tab:inputs}\\[-0.5em]}
	\begin{tabular}{rlrl}
      \toprule
	Parameter & Value & Parameter & Value \\
	\midrule
	$\alpha_s(M_Z)$ & $0.1179$ & 
    $\alpha_e(M_Z)$ & $1/127.952$ \\
    $\GF$ & $1.1663787\times 10^{-5}$ GeV$^{-2}$ &
    $M_Z$ & $91.1876$ GeV\\
    $M_h$ & $125.25$ GeV & 
    $m_e$ & $5.1099895$$\times 10^{-4}$ GeV \\
	$m_b(m_b)$ & $4.18$ GeV & 
    $m_c(m_c)$ & $1.27$ GeV\\
    \bottomrule
\end{tabular}
  \end{center}
\end{table}

The full expressions for
$F^{\text{(N)LL}}_\q$ are readily extracted from the results in
section~\ref{sec:rg}. For convenience, we give their numerical values
for the special case of $\muew=M_h$,
$\mu_b=m_b(m_b)$, and
$\mu_c=m_c(m_c)$. In this case, the logarithms in
$F^{\text{NLL}}_\q$ vanish and the only dependence is on
$\eta_{5/4}$.  We find
\begin{align}
  F_b^{\text{LL}}  &=-0.0202  \,,&
  F_b^{\text{NLL}} &=-0.0952  \,,\\ 
  F_c^{\text{LL}}  &=-0.383    \,,&
  F_c^{\text{NLL}} &=-0.685    \,,
\end{align}
where we used the values $\eta_5 = 0.506$ and
$\eta_4=0.605$ obtained by solving the mixed QCD--QED RG equations at
two-loop accuracy using the numerical input in Table~\ref{tab:inputs}.

\begin{figure}[h!]
\includegraphics[]{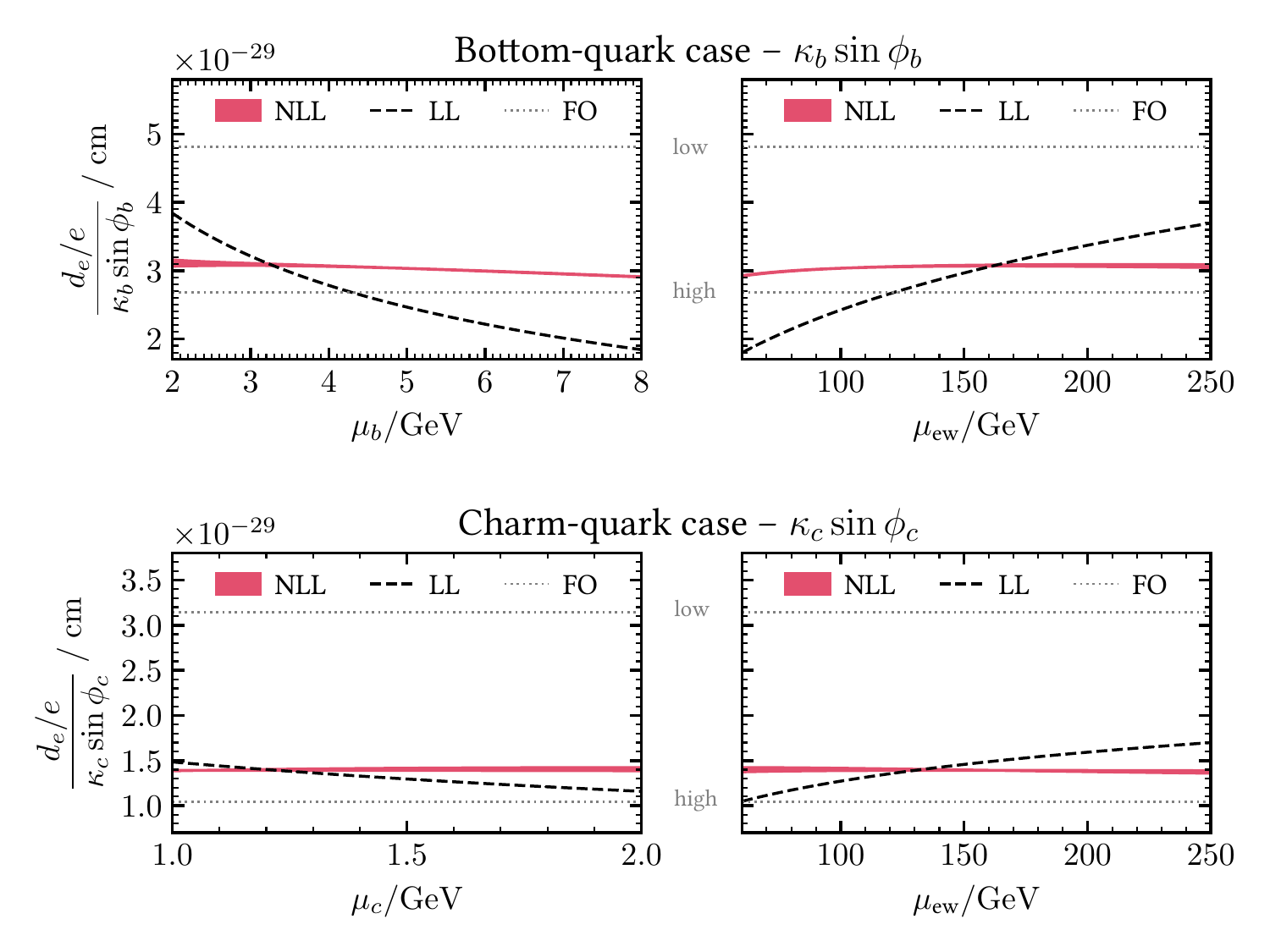}
\caption{Residual scale dependence of the electric dipole moment
  induced by CP-odd bottom-quark couplings (upper two panels) and
  CP-odd charm-quark couplings (lower two panels). In the left two
  panels, the variation of the matching scale at the bottom and charm
  thresholds is shown, respectively, while the right two panels show
  the dependence on the electroweak matching scale. In all plots, the
  dashed lines show the scale variation of the LL result. The scale
  variation of the NLL results is indicated by the red bands. The
  boundaries of the red bands are obtained by the evaluating the NLL
  scale variation in two ways, as discussed in the main text. Finally,
  the two gray dotted lines show the fixed-order results that have
  been used in literature so far; they correspond to evaluating the
  quark masses in Eq.~\eqref{eq:dsw} at the electroweak scale (the
  line marked ``high'') and at the heavy-quark threshold (the line
  marked ``low''), and neglecting all other QCD corrections.
  \label{fig:debottomcharm}}
\end{figure}

Next, we use the full analytic expressions to estimate the uncertainty
in predicting $d_e$, and provide the corresponding bounds on the
anomalous CP-odd $\q$-Yukawa couplings. We estimate the uncertainty
due to the truncation of the perturbation series in two ways:
\begin{itemize}
\item The dependence on the matching scales cancels in our result to
  the order we calculated ($\tilde \alpha_s \kappa^2$). The residual
  scale dependence is sensitive to higher-order terms in the
  perturbation series. Therefore, we evaluate the Wilson coefficient
  $\widetilde C^e_3$ at the fixed low scale $m_\q(m_\q)$, and
  separately vary all matching scales ($\muew$ and $\mu_b$ for the
  bottom-quark case; $\muew$, $\mu_b$, and $\mu_c$ for the charm-quark
  case). We fix all scales that are not varied to their ``central''
  values ($\muew = M_h$ and $\mu_\q=m_\q(m_\q)$). The maximal residual
  scale dependence then provides our first uncertainty estimate on
  $\widetilde{C}^e_3$ or, equivalently, on $d_e$.  The ranges for the
  scale variations are chosen as
  $\muew\in[60~\text{GeV},~250~\text{GeV}]$,
  $\mu_b=[2~\text{GeV},~8~\text{GeV}]$, and
  $\mu_c=[1~\text{GeV},~2~\text{GeV}]$. The scale variations are shown
  in Figure~\ref{fig:debottomcharm} and further discussed below. (The
  $\mu_b$ variation is not explicitly shown for the charm-quark case,
  as it looks very similar to the $\mu_c$ variation.)
\item There is a further ambiguity in our result that would only be
  resolved by a NNLL calculation: we can evaluate the NLL correction
  to $d_e$, i.e., the {\itshape whole} term proportional to
  $F^{\text{NLL}}_\q$ in Eq.~\eqref{eq:dresum}, using either two- or
  one-loop values for all masses and couplings. We use this numerical
  difference as a further way to estimate the uncertainty; this
  difference effectively smears the lines obtained for the NLL scale
  variation, as described above, into the red bands shown in
  Figure~\ref{fig:debottomcharm}.
\end{itemize}

As central values for $\widetilde{C}^e_3$ or equivalently $d_e$ at NLL
accuracy we take the average of the maximal and minimal values
obtained by all scale variations and differences as described
above. Half of that difference is then assigned as the theoretical
uncertainty associated with missing higher-order terms.  This leads to
  \begin{equation}
    \dfrac{d_e}{e} = {\kappa_\q \sin\phi_\q} \times 
    \begin{cases}
      (3.03\pm 0.13)\cdot 10^{-29}~\text{cm} \qquad\text{for}\quad \q=b\,,\\
      (1.39\pm 0.03)\cdot 10^{-29}~\text{cm} \qquad\text{for}\quad \q=c\,.
    \end{cases}
  \end{equation}

The corresponding results are further illustrated in the two upper and
two lower plots of Figure~\ref{fig:debottomcharm} for the bottom- and
charm-quark cases, respectively.  The plots on the left show the scale
dependence of $d_e/e$ on $\mu_\q$ at LL (dashed, black line) and NLL
(red band) accuracy.  The plots on the right show the corresponding
dependence on $\muew$. For the LL result we use one-loop values for
all masses and couplings, and both one- and two-loop values for the
NLL result, as discussed above. The comparison of LL and NLL
transparently shows how the NLL computation presented here drastically
reduces the large uncertainties associated to QCD corrections.

Furthermore, the gray dotted lines in Figure~\ref{fig:debottomcharm}
show the naive result of the fixed-order calculation,
Eq.~\eqref{eq:dsw}, that has been used so far in the literature. The
two lines correspond to using different values for the heavy-quark
mass $m_\q$: the line marked ``high'' corresponds to using the value
evaluated at the electroweak scale ($m_\q(M_h)$), while line marked
``low'' corresponds to using the mass evaluated at the low scale
($m_\q(m_\q)$). The spread of these three lines illustrates the level
of ambiguity in the fixed-order result. The NLL computation of the
current work removes this ambiguity almost entirely.

The electron EDM has been indirectly constrained by measuring the EDM
of polar molecules; the currently strongest bound is
$|d_e| < 4.1 \times 10^{-30}$ at 90\% confidence
level~\cite{Roussy:2022cmp}. To perform a rough combination of this
experimental constraint with our derived theory uncertainties we
interpret the measurement as a Gaussian centered at zero and the above
bound as the corresponding 90\% confidence level (CL) interval, i.e.,
including negative values for $d_e$. Adding the corresponding
experimental ``$1\sigma$'' uncertainty in quadrature with the theory
uncertainty we find based on an {\itshape one-parameter} $\chi^2$
function in terms of $\kappa_\q\sin\phi_\q$
\begin{align}
  \kappa_b|\sin\phi_b| &< 0.08 \quad\text{[at 68.27\% CL]}\,, & &\kappa_b|\sin\phi_b| < 0.14\quad\text{[at 90\% CL]}\,,\\
  \kappa_c|\sin\phi_c| &< 0.18 \quad\text{[at 68.27\% CL]}\,,  & &\kappa_c|\sin\phi_c| < 0.30\quad\text{[at 90\% CL]}\,.
\end{align}

\section{Discussion and conclusions\label{sec:conclusions}}

The experimental bound on the electron EDM~\cite{Roussy:2022cmp} (see
also Ref.~\cite{ACME2018}) translates into strong constraints on new
CP-violating phases in various extensions of the SM. This is
significant because the SM contribution to the electron EDM is
estimated to lie nine orders of magnitude below the current
experimental sensitivity~\cite{Pospelov:1991zt, Pospelov:2013sca}. In
this context, it is important to point out that the current strongest
bounds on the electron EDM are obtained indirectly via the measurement
of the EDM of polar molecules; it has recently been shown that the SM
contribution to those EDMs lies only about five orders of magnitudes
below the current experimental limit~\cite{Ema:2022yra}. Moreover, in
addition to the electron EDM operator, CP-odd semileptonic
four-fermion operators contribute to the EDM of polar molecules. While
such operators are generated in our model, their contributions are
suppressed by roughly five orders of magnitude compared to the
electron EDM, and can be safely neglected.

CP-violating phases such as those in Eq.~\eqref{eq:LagHqq} appear in
several well-motivated beyond standard model theories (see e.g.
Refs.~\cite{Feruglio:1992wf, Buchalla:2013rka}), and the electron EDM
is capable of placing stringent bounds on these phases; these have
mainly been studied in an EFT approach~\cite{Brod:2013cka,
  Chien:2015xha, Cirigliano:2016njn, Cirigliano:2016nyn,
  Cirigliano:2019vfc, Bahl:2022yrs, Brod:2022bww, Brod:2018pli}. As
shown in Ref.~\cite{Brod:2018pli}, analoguous, albeit weaker bounds on
CP-odd bottom and charm Yukawas can aso be obtained from hadronic
EDMs.

Implicit in the electron EDM bounds is a large ${\cal O}(1)$ QCD
uncertainty that has so far been neglected, even though it leads to
sizeable ambiguities in the resulting constraints.  Fortunately, since
the electron EDM is a leptonic observable, this ambiguity can be
removed systematically by a perturbative calculation, without the need
for additional non-perturbative information. In this work, we have
calculated the contribution to the electron EDM of CP-odd Higgs
couplings to the bottom and charm quarks in RG-improved perturbation
theory, summing the leading and next-to-leading large logarithms
proportional to the strong coupling constant. This calculation has
reduced the residual ambiguity in the bound to the level of a few
percent, as discussed in detail in Section~\ref{sec:numerics}. The
perturbation series shows good convergence, as expected for a leptonic
observable. If, in the future, a non-zero electron EDM were observed,
the error could be further reduced by summing the
next-to-next-to-leading QCD logarithms, as well as taking into account
the QED evolution of the electric dipole moment below the heavy-quark
thresholds.

\section*{Acknowledgments}

We thank Luca Merlo for a discussion that triggered this
project. Z.P.~acknowledges financial support from the European Research
Council (ERC) under the European Union's Horizon 2020 research and
innovation program under grant agreement 833280 (FLAY), and from the
Swiss National Science Foundation (SNF) under contract
200020-204428. J.B.~acknowledges support in part by DoE grant
DE-SC0011784.

\appendix
\section{Unphysical Operators\label{app:unphys}}

For completeness, we list here the ``unphysical'' operators entering
our calculation in intermediate steps. They are called ``unphysical''
because they vanish either via the equations of motion (e.o.m.)  of
the quark fields for onshell external states, or via algebraic
relations that are valid in $d=4$, but not in $d\neq 4$.

\subsection{E.o.m.-vanishing Operators}
These operators have matrix-elements that vanish via the e.o.m.\ of
the quark field. The following two gauge-invariant operators enter our
computation at the two-loop level:
\begin{equation} \label{eq:eom:LO}
\begin{split}
N_1^e
 & = \frac{m_\q}{2e^2} \bar e
     \bigl[ \stackrel{\leftarrow}{\slashed{D}}\stackrel{\leftarrow}{\slashed{D}} i\gamma_5
            + i\gamma_5 \slashed{D}\slashed{D}\bigr] e \,, \\
N_2^e
 & = \frac{m_\q}{2e^2} \bar e
     \bigl[   \stackrel{\leftarrow}{\slashed{D}} \stackrel{\leftarrow}{D^\sigma}
              \gamma^\mu\gamma^\nu\gamma^\rho
            - \gamma^\mu\gamma^\nu\gamma^\rho D^\sigma \slashed{D} \bigr] e \,\,
     \epsilon_{\mu\nu\rho\sigma}\,.
\end{split}
\end{equation}
%Additionally, the following operator, which is not gauge-invariant, is
%also required in intermediate steps to determine all counterterms and
%project the off-shell amplitudes
%\begin{equation} \label{eq:eom:LO:gauge-variant}
%N_{\gamma}^e
% = -\frac{im_b Q_e}{2e} \bar e
%      \bigg[\stackrel{\leftarrow}{\slashed{D}} \slashed{A} i \gamma_5
%            - i \gamma_5\slashed{A} \slashed{D} \bigg] e \,.
%\end{equation}
The covariant derivative acting on electron fields is defined as
\begin{equation}
D_\mu \equiv \partial_\mu + i e Q_e A_\mu \,,
\end{equation}
with $Q_e = -1$ the electron electrical charge.

\subsection{Evanescent Operators}
Next we list the evanescent operators that enter our computation at
one- and two-loop order.  The leading-order ADM does not depend on
their definition, but the next-to-leading-order ADM does.
In the $q$--$q$--$A$ sector we need the operator
\begin{equation} \label{eq:evan:qA:larin}
\begin{split}
  {E}_{\gamma }^{e} &= \frac{Q_e}{4} \frac{m_\q}{e}
            \bar e \{\sigma^{\mu\nu}, i \gamma_5\} e \, F_{\mu\nu} - Q_{3}^e\,.
\end{split}
\end{equation}
The evanescent operators required in the $e$--$\q$ sector read
\begin{equation} \label{eq:evan:eb}
\begin{split}
{E}_1^{e\q} &=\frac{1}{2}
(\bar e   \gamma_{[\mu}\gamma_{\nu]} e) \,
(\qb  \{\gamma^{[\mu}\gamma^{\nu]}, i\gamma_5\} \q)
                  + O_2^{e\q}\,,\\
{E}_1^{\q e} &=\frac{1}{2}
(\qb   \gamma_{[\mu}\gamma_{\nu]} \q) \,
(\bar e  \{\gamma^{[\mu}\gamma^{\nu]}, i\gamma_5\} e)
                  + O_2^{e\q}\,,\\
{E}_{2}^{e\q} &=
\frac{}{}
\big[
(\bar e  \gamma_{[\mu}\gamma_{\nu]}\gamma^{[\rho}\gamma^{\sigma]} e) \,
(\qb \gamma^{[\mu}\gamma^{\nu]}\gamma^{[\tau}\gamma^{\upsilon]} \q)
+
(\bar e  \gamma^{[\rho}\gamma^{\sigma]}\gamma_{[\mu}\gamma_{\nu]} e) \,
(\qb \gamma^{[\tau}\gamma^{\upsilon]}\gamma^{[\mu}\gamma^{\nu]} \q)
\big] \epsilon_{\rho\sigma\tau\upsilon}\\
&\quad
-48 (O_1^{e\q}+O_1^{\q e})
+16 O_2^{e\q}\,,\\
{E}_{3}^{e\q} &= \frac{1}{2}
(\bar e   \gamma_{[\mu}\gamma_{\nu}\gamma_{\rho}\gamma_{\sigma]} e) \,
(\qb  \{\gamma^{[\mu}\gamma^{\nu}\gamma^{\rho}\gamma^{\sigma]}, i\gamma_5\} \q)
                  -24 O_1^{\q e}\,,\\
{E}_{3}^{\q e} &= \frac{1}{2}
(\qb   \gamma_{[\mu}\gamma_{\nu}\gamma_{\rho}\gamma_{\sigma]} \q) \,
(\bar e  \{\gamma^{[\mu}\gamma^{\nu}\gamma^{\rho}\gamma^{\sigma]}, i\gamma_5\} e)
                  -24 O_1^{e\q}\,,\\
{E}_{4}^{e\q} &=\frac{1}{2}
(\bar e   \gamma_{[\mu}\gamma_{\nu}\gamma_{\rho}\gamma_{\sigma}\gamma_{\tau}\gamma_{\upsilon]} e) \,
(\qb  \{\gamma^{[\mu}\gamma^{\nu}\gamma^{\rho}\gamma^{\sigma}\gamma^{\tau}\gamma^{\upsilon]}, i\gamma_5\} \q) \,,\\
{E}_{4}^{\q e} &=\frac{1}{2}
(\qb   \gamma_{[\mu}\gamma_{\nu}\gamma_{\rho}\gamma_{\sigma}\gamma_{\tau}\gamma_{\upsilon]} \q) \,
(\bar e  \{\gamma^{[\mu}\gamma^{\nu}\gamma^{\rho}\gamma^{\sigma}\gamma^{\tau}\gamma^{\upsilon]}, i\gamma_5\} e) \,,\\
{E}_{5}^{e\q} &=
\frac{1}{2}
\big[
(\bar e  \gamma_{[\mu}\gamma_{\nu}\gamma_{\rho}\gamma_{\sigma]}\gamma^{[\tau }\gamma^{\upsilon]} e) \,
(\qb \gamma^{[\mu}\gamma^{\nu}\gamma^{\rho}\gamma^{\sigma]}\gamma^{[\zeta}\gamma^{\xi    ]} \q)\\
&\qquad +
(\bar e  \gamma^{[\tau }\gamma^{\upsilon]}\gamma_{[\mu}\gamma_{\nu}\gamma_{\rho}\gamma_{\sigma]} e) \,
(\qb \gamma^{[\zeta}\gamma^{\xi    ]}\gamma^{[\mu}\gamma^{\nu}\gamma^{\rho}\gamma^{\sigma]} \q)
\big]\epsilon_{\tau\upsilon\zeta\xi}
+ 48 O_2^{e\q} \,.
\end{split}
\end{equation}
The square brackets denote antisymmetrisation normalised as
\begin{equation*}
\gamma_{[\mu_1,...,\mu_n]} \equiv \frac{1}{n!}\sum_\sigma (-1)^\sigma \gamma_{\mu_{\sigma(1)}}\ldots \gamma_{\mu_{\sigma(n)}}\,.
\end{equation*}

\subsection{Operators Related to the Infrared Rearrangement}
The last class of unphysical operators arises because our use of infrared
rearrangement breaks gauge invariance in intermediate steps of the
calculation. At the renormalisable level this method generates two
gauge non-invariant operators corresponding to a gluon-mass and a
photon-mass term, i.e.
\begin{equation}
  \Lag \supset \frac{1}{2} Z_{\text{IRA},g} Z_G \, m_{\text{IRA}}^2 G_\mu^a G^{\mu,\,a}
               +  \frac{1}{2} Z_{\text{IRA},\gamma} Z_A \, m_{\text{IRA}}^2 A_\mu A^{\mu} \,.
\end{equation}
The mass, $m_{\text{IRA}}$, is completely artificial and
drops out of all physical results, and $Z_{\text{IRA},g}$,
$Z_{\text{IRA},\gamma}$ are two additional renormalisation
constants~\cite{Chetyrkin:1997fm}. One-loop insertions of the
dimension-five and dimension-six operators can induce further
gauge-invariant, higher dimension operators that are relics of the
infrared rearrangement. For our calculation, the only relevant one is
\begin{equation}
  \begin{split}
    P^e          &= m_\q \frac{m_{\text{IRA}}^2}{e^2} \bar e i \gamma_5 e \,.
  \end{split}
\end{equation}

\section{Renormalization Constants\label{app:ren_consts}}

The following are the SM counterterms necessary for the calculation:
\begin{align}
  Z_e^{(0,1)} & = \frac{2}{3\epsilon}
              \Big[ N_c \big( n_d Q_d^2 + n_u Q_u^2 \big) + n_\ell Q_e^2 \Big] \,, \\
  Z_e^{(0,2)} & = \frac{1}{\epsilon^2}
                 \bigg\{ \frac{2}{3}
                 \Big[ N_c \big( n_d Q_d^2 + n_u Q_u^2 \big) + n_\ell Q_e^2 \Big]^2
                 + \epsilon \Big[ n_\ell Q_e^4 + N_c \big(n_d Q_b^4 + n_u Q_c^4 \big) \Big]
                 \bigg\} \,, \\
  Z_A^{(0,1)} & = - \frac{4}{3\epsilon}
              \Big[ N_c \big( n_d Q_d^2 + n_u Q_u^2 \big) + n_\ell Q_e^2 \Big] \,, \\
  Z_A^{(1,1)} & = \frac{1}{\epsilon}
                 \big(1 - N_c^2 \big) \big( n_d Q_d^2 + n_u Q_u^2 \big) \,, \\
  Z_A^{(0,2)} & = - \frac{2}{\epsilon}
                 \bigg[ n_\ell Q_e^4 + N_c \big(n_d Q_d^4 + n_u Q_u^4 \big) \bigg] \,, \\
  Z_{m_A}^{(0,1)} & = - \frac{8}{3\epsilon}
              \Big[ N_c \big( n_d Q_d^2 + n_u Q_u^2 \big) + n_\ell Q_e^2 \Big] \,, \\
  Z_{\q}^{(1,0)} & = - \frac{1}{\epsilon} C_F \xi_g \,, \\
  Z_{f}^{(0,1)} & = - \frac{1}{\epsilon} Q_f^2 \xi_a \,, \\
  Z_{\q}^{(1,1)} & = \frac{1}{\epsilon^2}
                    C_F Q_\q^2 \bigg( \xi_g \xi_a + \frac{3\epsilon}{2} \bigg) \,, \\
  Z_{f}^{(0,2)} & = \frac{1}{\epsilon^2}
                    Q_f^2 \bigg\{ \frac{\xi_a^2}{2} Q_f^2
                    + \epsilon \bigg( \frac{3}{4} Q_f^2
                       + n_\ell Q_e^2
                       + N_c \big( n_d Q_d^2 + n_u Q_u^2 \big) \bigg) \bigg\} \,, \\
  Z_{m_\q}^{(1,0)} & = - \frac{3}{\epsilon} C_F \,, \\
  Z_{m_f}^{(0,1)} & = - \frac{3}{\epsilon} Q_f^2 \,, \\
  Z_{m_\q}^{(1,1)} & = \frac{9}{\epsilon^2}
                    C_F Q_\q^2 \bigg( 1 - \frac{\epsilon}{6} \bigg) \,, \\
  Z_{m_\q}^{(2,0)} & = - \frac{1}{\epsilon^2} \frac{C_F}{2N_c}
		    \bigg[\frac{1}{2} \bigg( 9 - 31 N_c^2 + 4 N_c (n_u + n_d) \bigg)
			  - \frac{\epsilon}{12} \bigg( 9 - 203N_c^2 + 20 N_c (n_u + n_d) \bigg) \bigg] \,, \\
  \begin{split}
  Z_{m_f}^{(0,2)} & = \frac{1}{\epsilon^2} Q_f^2
                    \bigg\{ \frac{9}{2} Q_f^2
                            - 2 \big( n_\ell Q_e^2
                                     + N_c n_d Q_d^2
                                     + N_c n_u Q_u^2 \big) \\
                & \hspace{6em} - \epsilon \bigg( \frac{3}{4} Q_f^2
                  - \frac{5}{3} \big(n_\ell Q_e^2
                             + N_c n_d Q_d^2
                             + N_c n_u Q_u^2 \big) \bigg) \bigg\} \,,
  \end{split}
\end{align}
where we have organised the contributions according to
\begin{equation}
	Z_i = \sum_{m,n}\tilde{\alpha}_s^m \tilde{\alpha}_e^n Z_i^{(m,n)}\,.
\end{equation}
and $f$ denotes either a charged-lepton or a quark field.

To obtain the two-loop anomalous dimension of the physical sector we
need certain one-loop renormalisation constants involving unphysical
operators. We collect them in this appendix.  
We write $Z_{x\to y}$
where the subscripts $x$ and $y$ symbolize sets of Wilson
coefficients, for which we use the following notation and standard
ordering:
\begin{equation}
	\begin{split}
P = \{
&C_1^{e\q},\, C_1^{\q e},\, C_2^{e\q},\, C_3^{e}\}\,, \\
E = \{
  & C_{E_1^{e\q}},\,
    C_{E_1^{\q e}},\, C_{E_2^{e\q}},\,
    C_{E_{\gamma }^{e}}\}\,, \\
M = \{ &C_{P^{e}}\}\,,\\
N = \{  &C_{N_1^{e}},\,C_{N_2^{e}}\}\,.
	\end{split}
\end{equation}

The first necessary input is the mixing of the physical operators into
all the evanescent operators that are generated at one-loop. Using the
same subscript notation as above, the renormalisation constants read
\begin{equation}
Z^{(0,1)}_{P\to E} =
\begin{pmatrix}
Q_e Q_\q& 0& 0& 0\\
 0&Q_e Q_\q& 0& 0\\
 0& 0& - \frac{1}{2} Q_e Q_\q& 0\\
 0& 0& 0& -4Q_e^2
\end{pmatrix}\,,
\end{equation}
and $Z^{(1,0)}_{P\to E} = 0$. The remaining mixing of physical
operators into evanescent operators is zero at one-loop. Furthermore,
the finite part of the mixing of evanescent into physical operators is
subtracted by finite counterterms~\cite{Dugan:1990df}. They read
\begin{equation}
Z^{(0,1)}_{E\to P} =
\begin{pmatrix}
4 Q_e Q_\q& 0& 0& 0\\
 0&4 Q_e Q_\q& 0& 0\\
 0& 0& - 8 Q_e Q_\q& 0\\
 0& 0& 0& -\frac{2}{3}Q_e^2
\end{pmatrix}\,.
\end{equation}
The remaining finite mixing of evanescent into physical operators is
zero at one-loop.

Furthermore, we need the mixing constants of the physical operators
into the operators arising from infrared rearrangement; they are found
to be
\begin{equation}
Z^{(0,1)}_{P\to M} =
\begin{pmatrix}
0\\
-12\\
0\\
-12 Q_e^2
\end{pmatrix}
\,.
\end{equation}
All other mixing constants of physical into the IRA
operators are zero.

Finally, we need the mixing constants of the physical operators into
the e.o.m.-vanishing operators. They are uniquely fixed by the
$\q\to\q$ Greens function. We find
\begin{equation}
Z^{(0,1)}_{P\to N} =
\begin{pmatrix}
0&0\\
0&0\\
0&0\\
-2Q_e^2&-\frac{1}{6}Q_e^2
\end{pmatrix}\,.
\end{equation}

The two-loop anomalous-dimension matrix is given in terms of the one-
and two-loop renormalisation constants by
\begin{align}
\gamma^{(0,2)} & = 4 Z^{(0,2;1)} - 2 Z^{(0,1;1)} Z^{(0,1;0)} \,, \\
\gamma^{(1,1)} & = 4 Z^{(1,1;1)} - 2 Z^{(0,1;1)} Z^{(1,0;0)}
                  - 2 Z^{(1,0;1)} Z^{(0,1;0)} \,, \\
\gamma^{(2,0)} & = 4 Z^{(2,0;1)} - 2 Z^{(1,0;1)} Z^{(1,0;0)} \,,
\end{align}
where we have further separated the renormalisation constants according to their
poles, i.e.
\begin{equation}
	Z_i^{(m,n)} = \sum_r \frac{1}{\epsilon^r}Z_i^{(m,n;r)}
\end{equation}
The quadratic poles of the two-loop diagrams are fixed by the poles of
the one-loop diagrams via
\begin{equation}\label{eq:2loop:rel}
  \begin{split}
Z^{(0,2;2)} &= \frac{1}{2} Z^{(0,1;1)} Z^{(0,1;1)} - \frac{1}{2} \beta_e Z^{(0,1;1)} \,, \\
Z^{(1,1;2)} &= \frac{1}{2} Z^{(0,1;1)} Z^{(1,0;1)} + \frac{1}{2} Z^{(1,0;1)} Z^{(0,1;1)} \,, \\
Z^{(2,0;2)} &= \frac{1}{2} Z^{(1,0;1)} Z^{(1,0;1)} - \frac{1}{2} \beta_0 Z^{(1,0;1)} \,,
  \end{split}
\end{equation}
where $\beta_0=\frac{11}{3}N_c-\frac{2}{3}\Nf$. As a check of our
calculation, we computed these poles directly and verified that they
satisfy Eq.~\eqref{eq:2loop:rel}.

\addcontentsline{toc}{section}{References}
\bibliographystyle{JHEP}
\bibliography{references}

\end{document}